\documentclass[submission,copyright,creativecommons]{eptcs}
\providecommand{\event}{HCVS 2025} 
\usepackage{t1enc}
\usepackage{microtype}
\usepackage{alltt}

\usepackage{underscore}         
\usepackage[T1]{fontenc}        

\usepackage{latexsym}
\usepackage[all]{xy}
\usepackage{graphicx}

\usepackage{amsmath}
\usepackage{amssymb}
\usepackage{stmaryrd}
\usepackage{moreverb,epsfig,url}
\input{prolog-highlighting.tex}

\newcommand{\shortImplique}{\:\Rightarrow\:}

\newcommand{\rw}{\rightarrow}

\newfont{\amstoto}{msbm10}
\newcommand{\NN}{\mbox{\amstoto\char'116}}

\newcommand{\mydef}{\mathrel = \!\!\!\!\!\!\!\raisebox{2pt}{$^{def}$}}

\title{Finding Regular Herbrand Models for CHCs \\  using Answer Set Programming}

\author{Grégoire Maire \institute{ENS Rennes}
  \email{gregoire.maire@ens-rennes.fr} \and
Thomas Genet \institute{Univ Rennes, IRISA, Inria} \email{genet@irisa.fr}}

\def\titlerunning{Finding Regular Herbrand Models for CHCs using Answer Set Programming}
\def\authorrunning{\textsc{G.~Maire, T.~Genet}}

\begin{document}

\maketitle

\begin{abstract}
  We are interested in proving satisfiability of Constrained Horn Clauses (CHCs)
  over Algebraic Data Types (ADTs). We propose to prove satisfiability by
  building a tree automaton recognizing the Herbrand model of the CHCs. If such
  an automaton exists then the model is said to be {\em regular}, i.e., the
  Herbrand model is a regular set of atoms.
  Kostyukov \& al.~\cite{kostyukovmf-pldi21} have shown how to derive an
  automaton when CVC4 finds a finite model of the CHCs.
  We propose an alternative way to build the automaton using an encoding into a
  SAT problem using Clingo, an Answer Set Programming (ASP) tool.
  We implemented a translation of CHCs with ADTs
  into an ASP problem. Combined with Clingo, we obtain a semi-complete
  satisfiability checker: it finds a tree automaton if a regular Herbrand model
  exists or finds a counter-example if the problem is unsatisfiable.
\end{abstract}

We are interested in the automatic verification of programs manipulating
Algebraic Data Types (ADTs). The analysis of such programs is challenging
as soon as the ADTs are recursive because they define unbounded data
structures. When ADTs are recursive, tree automata~\cite{TATA} provide an
efficient way to finitely represent unbounded sets of such
ADTs. In~\cite{MatsumotoKU-APLAS15,GenetHJ-FOSSACS18,HaudebourgGJ-ICFP20,kostyukovmf-pldi21}, 
verifying a property $\phi$ on a program $P$ consists in building a tree
automaton recognizing a set of all the computations of the program and in
checking that the property is true on this set. When $P$ is represented by a set
of functions (resp. a term rewriting system), the property $\phi$ is expressed
as a set of results that should not be reachable when applying the semantics of
$P$ on initial function calls (resp. rewriting initial terms with $P$). In this
setting the tree automaton finitely represents the set of all values
(resp. terms) that are reachable when applying the semantics of $P$
(resp. rewriting with $P$).
In the context of program verification using
Constrained Horn Clauses (CHC for short), this is adapted as follows: $P$ is
represented by a set of Horn clauses and the property $\phi$ is a negative
formula, i.e., $\phi\mydef (\psi \shortImplique \bot)$. To prove
that $P \shortImplique (\psi \shortImplique \bot)$ is valid, we build a tree
automaton finitely representing the least Herbrand model $M$ 
of $P$ and we check that the formula $\psi$ does not hold in $M$. This
entails that $\psi$ is {\em not} a logical consequence of $P$.
Thus, $P \shortImplique (\psi \shortImplique \bot)$ is valid. In the following,
if a model can be represented by a tree automaton we call it a {\em regular
  model}, i.e., the
  Herbrand model is a regular set of atoms.

\section{An introductory example}
\label{example}
For instance, let $nat = z~|~s(nat)$ be the ADT defining natural numbers. Let
$P$ be the set of CHCs defining $even(x)$ and $odd(x)$ as the usual predicates
over numbers and $plus(x,y,z)$ as the predicate such that $z= x + y$. Let $\phi_1
\mydef ~\forall x~y~z. ~ even(x) \wedge even(y) \wedge plus(x,y,z) \wedge odd(z)
\shortImplique \bot$ be the (negative) property we want to prove on $P$.
Since the ADT of natural numbers is recursive, Herbrand models of $P$ are
unbounded. However, we can finitely represent such an unbounded model using a
tree automaton: 

$\begin{array}{r@{\hspace{2cm}}r@{\hspace{2cm}}r}
  z \rw \#2       & odd(\#1) \rw \#0           & plus(\#1,\#2,\#1) \rw \#0 \\
s(\#2) \rw \#1  & even(\#2) \rw \#0          &   plus(\#1,\#1,\#2) \rw \#0 \\
s(\#1) \rw \#2  & plus(\#2,\#1,\#1) \rw \#0  &   plus(\#2, \#2, \#2) \rw \#0
\end{array}$

\noindent
In this automaton, \#0, \#1, \#2 are the states of the automaton. A term is recognized by
a state if it can be rewritten to this state using the transitions. For instance,
the state \#2 recognizes the term $z$, the state \#1 recognizes $s(z) \rw s(\#2)
\rw \#1$, i.e., \#2 recognizes even numbers and \#1 recognizes odd
numbers. Finally, the state $\#0$ recognizes all the atoms that are true in the
considered Herbrand model, e.g, $even(s(s(z)))$, $odd(s(z))$. Note that
the model recognized by this automaton is not the least Herbrand model but an
over-approximation. In particular, this automaton recognizes 
$plus(s(z),s(z),s(s(z)))$ which is part of the least Herbrand model but also
$plus(s(z),s(z),z)$ which is not. On this model, the property $\phi_1$ is
true. In particular, with the rule $plus(\#2, \#2, \#2) \rw \#0$ we can see that
summing two even numbers always result into an even number. Since the negative
property $\phi_1$ is true on an over-approximation of the least Herbrand model
then it is also true on the least Herbrand model.

To infer such an automaton, the tool RInGen by Kostyukov \& al.~\cite{kostyukovmf-pldi21} use the
finite model finder of CVC4. They transform the input problem $(P \wedge \phi)$ over the theory of
ADTs into a problem in the Equality Logic with Uninterpreted Functions
(EUF). Then, if there exists a finite model of the problem in EUF, they show how
to derive a tree automaton recognizing an Herbrand model in the ADT theory
satisfying $P \wedge \phi$ and proving $P \shortImplique \phi$.

\section{Building automata recognizing regular models using ASP}
In this paper, we report preliminary experiments on an alternative way to
build such an automaton by an encoding into a SAT problem using
Clingo~\cite{clingo}, an Answer Set Programming (ASP) tool. Complex automata
inference with Clingo has already been experimented in~\cite{LosekootGJ-FSCD23,LosekootGJ-SAS24}.
Given a set of Prolog-style clauses, Clingo searches for a
Herbrand model of this set of clauses. Unlike usual Prolog interpreter, Clingo
is guaranteed to terminate and outputs the Herbrand models as soon as the models are
finite. However, how discussed above, the Herbrand models we
look for are {\em not} finite but can be {\em finitely} represented using tree
automata. Here is a possible encoding of $P$ and $\phi_1$ in Clingo.
We first set the maximal number of states in the automaton we search for.
\begin{lstlisting}[style= Prolog-pygsty]
#const maxState=2.      
state(1..maxState). % this shortcut builds facts state(1). and state(2).
\end{lstlisting}
The following lines define the tree automaton rules for the abstraction of the
ADT. We encode a rule of the form $s(\#1) \rw \#2$ by the fact 
\lstinline[style = Prolog-pygsty]|rule(s(1),2)|
The automaton we want to build for terms of
the ADT is expected to be complete (any term should be recognized by {\em at
  least} one state) and deterministic (any term should be recognized by {\em at
  most} one state). This is easily encoded using
Clingo's cardinality constraints.
\begin{lstlisting}[style= Prolog-pygsty]
1 {rule(z, Q): state(Q)} 1.
1 {rule(s(Q0), Q): state(Q)} 1 :- state(Q0).
\end{lstlisting}
  In the first line above, the brackets around the fact
  \lstinline[style = Prolog-pygsty]|rule(z,Q)|
  mean that this fact may or may not appear in the
searched model. By adding $1$ on the left, we impose that {\em at least} one
fact of this kind appears in the model. By adding $1$ on the right we impose
that {\em at most} one fact of this form appears in the model. The annotation
\lstinline[style = Prolog-pygsty]|state(Q)| forces {\tt Q} to be one of the states. Thus, if a model is found
it will necessarily have exactly one fact
\lstinline[style = Prolog-pygsty]|rule(z,1)| or
\lstinline[style = Prolog-pygsty]|rule(z,2)| 
(since we have here only 2 states). The second
line ensures there is exactly one state $Q$ such that $s(Q_0) \rw Q$ for all
states $Q_0$. The following lines essentially give the types and cardinality of
the relations $even$, $odd$ and $plus$. We provide those information but we want
to infer the relation themselves. Again, because of the
brackets, these lines only say that those facts may or may not appear in the
model.
\begin{lstlisting}[style= Prolog-pygsty]
{even(Q0)} :- state(Q0).
{odd(Q0)} :- state(Q0).
{plus(Q0, Q1, Q2)} :- state(Q0), state(Q1), state(Q2).
\end{lstlisting}
Finally, we can state the CHCs of our satisfiability problem. They are directly
translated into Clingo clauses where terms are replaced by the corresponding
states and transitions. For instance, one clause defining the $even$ predicate
is $even(s(X))~:-~ odd(X)$. In the encoding, since predicates
ranges over states and not terms, we cannot directly represent an atom over the
term $s(X)$. Instead, we encode this using several facts, i.e., a state $Q_1$
and a rule $s(Q_0) \rw Q_1$. This results into the following set of Clingo
clauses.

\medskip

\begin{lstlisting}[style= Prolog-pygsty]
% Translation of :    even(z).
even(Q0) :- rule(z, Q0).
% Translation of :    even(s(X)) :- odd(X).
even(Q2) :- odd(Q1), rule(s(Q1), Q2).
% Translation of :    odd(s(X)) :- even(X).
odd(Q2) :- even(Q1), rule(s(Q1), Q2).
% Translation of :    plus(z, X, X).
plus(Q1, Q0, Q0) :- rule(z, Q1), state(Q0).
% Translation of :    plus(s(X), Y, s(Z)) :- plus(X, Y, Z).
plus(Q6, Q3, Q7) :- plus(Q1, Q3, Q5), rule(s(Q1), Q6), rule(s(Q5), Q7).
% Translation of :    :- even(X), even(Y), plus(X, Y, Z), odd(Z).
:- even(Q0), even(Q1), plus(Q0, Q1, Q2), odd(Q2).
\end{lstlisting}

\medskip
\noindent
We prototyped this translation and the satisfiability checking in OCaml and
Clingo: \url{https://gitlab.inria.fr/regular-pv/regularmodels}. The translation is very close to the one above except that
it also uses a predicate
\lstinline[style = Prolog-pygsty]|stateType(Q,t)|
to distinguish
states w.r.t. the type
\lstinline[style = Prolog-pygsty]|t|
of the terms they recognize.
Another difference is that bodies of initial CHCs may contain equalities $X = Y$ or 
disequalities $X != Y$ ranging over terms. Equalities can be encoded by
equalities on states because the automaton is deterministic: if terms are equal
then so are the states. However, note that different terms may be
recognized by the same state. Hence, the body of the clause may be true on
states though it is not on the recognized terms. This results into an
over-approximation of the Herbrand model which is safe w.r.t. the property that
is a negative clause. On the opposite, encoding term disequalities by state
disequalities is not safe: a disequality $X !=Y$ in the body may be satisfied by
two different terms $t_1$ and $t_2$ though they are recognized by the same
state. This would result into an under-approximation of the Herbrand model which
is not safe. As a result we define the
\lstinline[style = Prolog-pygsty]|diffApprox(Q1,Q2)|
predicate over-approximating the $!=$ relation on terms.
This predicate is true if $Q1$ and $Q2$ recognizes at least two different terms.  

Finally, our satisfiability procedure generates Clingo specifications with
increasing values of
\lstinline[style = Prolog-pygsty]|maxStates|
until one solution is
found. For each value of
\lstinline[style = Prolog-pygsty]|maxStates|,
we generate two
specifications: one for satisfiability checking and another (with small
modifications) to search for a counterexample. Note that, given a value of
\lstinline[style = Prolog-pygsty]|maxStates|,
if Clingo fails to find a model (and if Clingo
is complete) then we have a guarantee that there exists no regular Herbrand
model that can be recognized by an automaton of
\lstinline[style = Prolog-pygsty]|maxStates| states.
Here is the output of our prototype on the example of Section~\ref{example}.

{\small
\begin{alltt}
Searching for a counterexample with 1 state
Searching for a model with 1 state
Searching for a counterexample with 2 states
Searching for a model with 2 states
ADT Transitions:        Predicates:
Z -> 2                  odd(1)             plus(1,2,1)
S(2) -> 1               even(2)            plus(1,1,2)
S(1) -> 2               plus(2,1,1)        plus(2,2,2)

Success! Clauses are satisfiable by a Herbrand model recognized by a tree 
automaton with 2 states
\end{alltt}}

Note that in the tree automaton of Section~\ref{example}, we also generated a
state (\#0) and transitions (e.g. $plus(\#1,\#2,\#1) \rw \#0$) to recognize the
terms rooted by predicate symbols. However those transitions are useless for
verification of CHCs and are, thus, discarded in the output of our prototype.
By iteratively increasing
\lstinline[style = Prolog-pygsty]|maxStates|,
we have a
semi-complete tool to check for satisfiability of CHCs with ADTs: if there
exists a regular Herbrand model we will find it. This was also the case
with RInGen~\cite{kostyukovmf-pldi21} where semi-completeness relies on completeness of
CVC4 finite model-finder.

\section{Experimental evaluation}
With regards to efficiency, our prototype is not yet as efficient
as RInGen. This is essentially due to the fact that our
Clingo encoding is too general: each Clingo specification may have several
equivalent solutions, i.e., several equivalent Herbrand models. Since efficiency
of the Clingo solving highly depends on the number of possible solutions, we
need to reduce the number of solutions to improve the efficiency of our
prototype. 
For instance, with the Clingo specification of the previous section, encoding
\lstinline[style = Prolog-pygsty]|plus(X,Y,Z)|,
\lstinline[style = Prolog-pygsty]|even(X)|, and
\lstinline[style = Prolog-pygsty]|odd(X)|, 
there are two
equivalent solutions. The solution automaton presented in the above section
recognizes odd numbers in state {\tt 1} and even numbers in state {\tt 2}. However, the
generated Clingo specification has a second equivalent and symmetrical
solution where odd numbers are recognized in state {\tt 2} and even numbers in
state {\tt 1}.

We studied the impact symmetries on Clingo's solving efficiency using
a more complex verification problem using two ADTs: the type
$elt$ of elements and the type $list$ of lists of $elt$. The ADT $elt$ contains
a finite set of $k$ constants where $k\in 
\NN$, i.e., $elt = a_1 | \ldots | a_k$. The $list$ ADT is defined by $list = nil
~|~ cons(elt,list)$. Let $P$ be the set of CHCs defining $member(x,l)$ as the
predicate which is true if the element $x$ belongs to the list $l$,
$notMember(x,l)$ as the negation of $member(x,l)$, and
$rev(l_1,l_2)$ such that $l_2$ is $l_1$ reversed. Assume that we want to prove
the property that an element belongs to a list if and only if it belongs to the reverse of
this list. This property can be encoded by the following two negative formulas $phi_2 \mydef ~\forall
x~l_1~l_2.~ member(x,l_1) \wedge reverse(l_1,l_2) \wedge notMember(x,l_2)
\shortImplique \bot$
and $phi_3 \mydef ~\forall
x~l_1~l_2.~ notMember(x,l_1) \wedge reverse(l_1,l_2) \wedge  member(x,l_2)
\shortImplique \bot$.

Having an algebraic data-type $elt$ whose size $k$ vary makes it possible to
increase the complexity of the verification problem by increasing $k$. We tried to prove the above verification
problem ($P \shortImplique \phi_2 \wedge \phi_3$)
for values of $k$ ranging from~$2$ to~$4$. We experimented with RInGen and our
prototype. For $k=2$, RInGen solves it in 0.075s while our tool solves it in
0.395s. For $k=3$, RInGen solves it in 1.211s while our tool solves it in
2700s (45 minutes!). This example shows that a naive ASP-encoding will fail
to efficiently build regular models. The influence of symmetries can be observed by
asking Clingo to generate the number of solutions for a given input
specification. With $k=2$ the number of solutions is greater than 700 
millions. With $k=3$ the number of solutions is so huge that Clingo fails to 
output it.

We modified by hand the Clingo specifications generated by our tool in order to
apply some simple symmetry breaking techniques. The objective is to find an
order on states that is restrictive enough to discard equivalent solutions and
permissive enough not to loose any valid solution. We applied this to the above
verification problem for values of $k$ from~$2$ to~$4$. We sum-up all those
experiments in the table Figure~\ref{fig:table}, where we compare the execution time for
RInGen (using CVC4 as a backend), the execution time for our ASP-prototype, the
number of equivalent models for our ASP-prototype, the execution time for our
ASP-prototype with symmetry breaking modification done by hand, and finally the 
corresponding number of models with symmetry breaking.

\begin{figure}[!ht]

\noindent
    \begin{tabular}{|@{~}c@{~}|l||l|l||l|l|}
      \hline
      $k=|{\color{red}elt}|$ &RInGen    & ASP-prot.      & ASP-prot.       & ASP-prot. sym. break.  & ASP-prot. sym. break.\\
              & CVC4 (sec.) & (sec.)      & \# models & (sec.)              & \#
                                                                              models \\
      \hline
      {\color{red}2}       & 0.075    & 0.395    & 700 M+ & 0.035            & 12  \\ 
      \hline
      {\color{red}3}       & 1.211    &
                                                                    2700
                                                 & Timeout      & 0.616           & Timeout \\
      \hline
      {\color{red}4}      & Timeout       & Timeout       & Timeout      & Timeout               & Timeout \\
      \hline
\end{tabular}
\caption{Experiments with RInGen, our prototype and our prototype with symmetry breaking\label{fig:table}}
\end{figure}

In this table, we can remark on line $k=2$ that even a simple symmetry breaking
dramatically reduce the number of considered models. The effect on
efficiency is valuable for $k=2$ but is really significant for $k=3$ where the
computation time decreases from 2700s to 0.616s. We even get an
execution time that is lower than the one of RInGen. However, our symmetry
breaking can still be improved since the number of models for $k=3$ is still too big
to be outputted by Clingo. Finally, the last remark
is that no implementation can solve this problem for $k=4$ and, thus, there is
still room for improvements!

We believe that the basic symmetry breaking we carried out by hand is correct, i.e.,
that it does not jeopardize the semi-completeness of the approach. However, this
has to be proven. If correct, our symmetry breaking strategy has to be integrated in our
prototype. The proof and implementation are left for future work. 
Besides, we believe that using an even more aggressive symmetry breaking
technique could yield a regular model finder more efficient than RInGen because
Clingo's solving core is pure SAT-solving. This has to be investigated
further. Another way to improve efficiency is to use a modular solving based on the Regular 
Language Typing approach of~\cite{HaudebourgGJ-ICFP20}. Finally, moving automata
inference from CVC4 to ASP-based solvers should open ways to infer automata with
constraints, e.g., tree automata with arithmetic constraints to verify programs
with ADTs containing numerical values. 

\vspace*{-4mm}
\paragraph*{Acknowledgments}

\noindent
Many thanks to Théo Losekoot and Jacques Nicolas for fruitful discussions about Clingo.

\bibliographystyle{eptcs}
\bibliography{genet}

\end{document}